\documentclass[prb,aps,epsf,twocolumn,showpacs]{revtex4}
\usepackage[dvips]{graphicx}
\usepackage{latexsym}
\usepackage{amsmath}
\begin{document}

\title{Coherence and pairing in a doped Mott insulator: Application to the cuprates}

\author{T. Senthil and Patrick A. Lee}
\affiliation{ Department of Physics, Massachusetts Institute of
Technology, Cambridge, Massachusetts 02139}

\date{\today}
\begin{abstract}
The issues of single particle coherence and its interplay with singlet pairing are studied within the slave boson gauge theory of a doped Mott insulator. Prior work by one of us\cite{mottcrit}  showed that the coherence scale below which Landau quasiparticles emerge is parametrically lower than that identified in the slave boson mean field theory. Here we study the resulting new non-fermi liquid intermediate temperature regime characterized by a single particle scattering rate that is linear in temperature ($T$). In the presence of a $d$-wave pair amplitude this leads to a pseudogap state with $T$ dependent Fermi arcs near the nodal direction. Implications for understanding the cuprates are discussed.

\end{abstract}
\newcommand{\be}{\begin{equation}}
\newcommand{\ee}{\end{equation}}
\newcommand{\bea}{\begin{eqnarray}}
\newcommand{\eea}{\end{eqnarray}}
\newcommand{\bK}{\textbf{K}}
\newcommand{\bp}{\textbf{p}}

\maketitle

In a recent paper\cite{tsleehitcphen} we developed a unified phenomenological picture of some of the most striking properties of the underdoped cuprates.
A key aspect of the phenomenology is the existence of a low `coherence' scale $T_{coh}$ for the single particle excitations below which Landau quasiparticles become well defined. We took $T_{coh}$ to rise as the doping $x$ is increased (roughly order the superconducting transition temperature $T_c$ in zero magnetic field), and to be only weakly affected by moderate field strengths. The purpose of this paper is to explore a specific microscopic model where the issue of electron coherence and its interplay with the singlet pairing (that undoubtedly also exists in the underdoped cuprates) can be studied in detail.

\begin{figure}
\includegraphics[width=8cm]{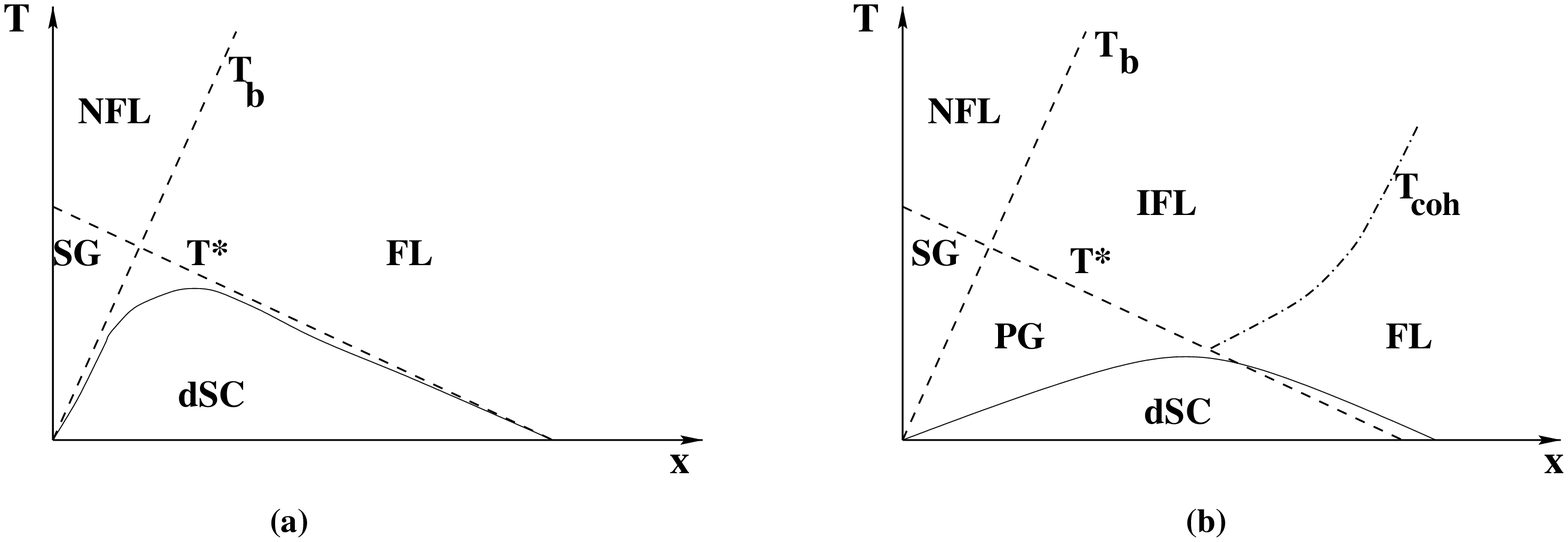}
\caption{Schematic slave boson phase diagram of the doped $t-J$ model. (a) The mean field phase diagram
(b) Phase diagram beyond mean field. All dashed lines represent crossovers.}
 \label{tJslvpdia}
\end{figure}

It has long been recognized that the properties of the cuprates must be understood within the framework of doping a Mott insulator\cite{lnwrmp}.
A useful theoretical framework  to discuss the doped Mott insulator is provided by a slave particle treatment of the ``$t-J$" model. In the standard ``slave boson" theory of the $t-J$ model\cite{kotliarliu,lnwrmp}, the mean field phase diagram looks as shown in Fig. \ref{tJslvpdia}a.  The slave boson field $b$ condenses at a temperature $T_b \sim T_c \sim x$ which is then identified with $T_{coh}$.  In addition $d$-wave singlet pairing of the spins sets in at a $T^*$ line that decreases with increasing doping. Fluctuations effects beyond the mean field theory have also been extensively addressed. At temperatures $T > T_b$, non-fermi liquid physics has been found\cite{ioffe,leenag}.

In this paper we revisit the slave particle gauge theory and show, following recent work by one of us\cite{mottcrit}, that true single particle coherence is not established till an energy scale $T_{coh}$ that is parametrically lower than the mean field estimate $T_b$. The existence of an intermediate temperature non-fermi liquid regime in between $T_b$ and $T_{coh}$
was pointed out relatively recently\cite{mottcrit}, and was missed in the prior literature. Ref. \onlinecite{mottcrit} briefly described some properties of this regime which we dub an ``Incoherent Fermi Liquid" (IFL). Here we study it in greater detail in the specific context of the cuprates. In the absence of pairing ({\em i.e} above the $T^*$ line), the IFL is characterized by a $T^{2/3}$ specific heat, a constant spin susceptibility, and a sharp Fermi surface leading to enhanced ``$2K_F$" response in both spin and charge channels. Most interestingly we show that  the single particle spectrum possesses peaks at the Fermi surface with a scattering rate $ \propto T$ unlike in a Landau fermi liquid. The crossover to the true Fermi liquid happens only at $T_{coh} \sim x^{\frac{3}{2}} \ll T_b$.
Below $T^*$ in the regime dubbed PG (for pseudogap) in Fig. \ref{tJslvpdia}b, this scattering persists despite the development of a $d$-wave pair amplitude  and leads to a pseudogap and $T$-dependent Fermi arc behavior.

All the prior theoretical work\cite{lnwrmp} on the strange metal and pseudogap phases within the slave boson framework has focused on the non-fermi liquid(NFL) and spin gap(SG) regimes of Fig. \ref{tJslvpdia}b. IFL or its descendant PG have not been studied thus far but may be more relevant to the temperatures actually probed in experiments. Indeed it was recognized early on\cite{leenag} that the slave boson mean field theory predicts a $T_b$ (and hence a coherence temperature) that is far too high (order $1000 K$ or higher at optimal doping).
Thus the lowering of the true coherence scale described above from the slave boson mean field theory is potentially
significant, and  the IFL, PG regimes need to be explored.

To describe our results, first consider the phase diagram in the absence of pairing, {\em i.e} above $T^*$. Then the system can be viewed as resulting from doping a spin liquid Mott insulator with  a Fermi surface of neutral spin-$1/2$ fermionic spinons $f_\alpha$ ($\alpha = \uparrow,\downarrow$). The physical electron operator $c_{\alpha} = b^\dagger f_{\alpha}$. Both $b$ and $f_{\alpha}$ are coupled to a $U(1)$ gauge field $a_\mu$.
Near the chemical potential tuned Mott transition, the slave boson condensation scale $T_b$ is of order the ground state boson phase stiffness $\rho_{bs}$. Naively once $b$ condenses, by the Anderson-Higgs mechanism the gauge fields are gapped out, and the $f_\alpha$ acquire electric charge to become Landau quasiparticles. The key point of Ref. \onlinecite{mottcrit} is that, in contrast to this naive expectation, the Anderson `plasmonization' of the gauge field occurs only at a scale $\sim \rho_{bs}^{3/2}$ which is much smaller than the `Higgs' scale $\rho_{bs}$ of the $b$ condensation. This is because of the unusual dynamics of the gauge field due to Landau damping by the fermions. At energy scales above the Anderson scale but below the Higgs scale the bosons are condensed but the gauge field can still be treated as gapless. This leads to strong scattering of the fermions and to non-fermi liquid physics.

We begin by deriving an effective field theory for energy scales below the scale $T_b$. In the absence of spinon pairing the full slave boson action (in imaginary time $\tau$) reads
\bea
S & = & S[f_\alpha, a] + S[b,a] + S[a] \\
S[f,a] & = & \int d\tau \sum_r \bar{f}_{r\alpha} \left(\partial_\tau - ia_0 - \mu_f \right) f_{r\alpha} \\
& & - \int_{\tau} \sum_{<rr'>} t_s \left(\bar{f}_{r\alpha}f_{r'\alpha} e^{ia_{rr'}} + h.c \right) \\
S[b,a] & = & \int_{\tau} \bar{b}_r \left(\partial_\tau - ia_0 - \mu_b \right)b_r \\
& & -\int_\tau t_c \left(\bar{b}_{r}b_{r'} e^{ia_{rr'}} + h.c \right) + V[\bar{b}_r b_r]
\eea
Here $r,r'$ are the sites of the square lattice. $\mu_b, \mu_f$ are the boson and fermion chemical potentials respectively. They are adjusted so that the condition $<\bar b_r b_r >  =  x, <\bar{f}_r f_r>  =  1-x$ are both satisfied.
$t_s,t_c$ are the hopping amplitudes of the fermion and boson respectively. The $V$ term represents a short range boson repulsion. We have not explicitly written out the gauge action $S[a]$ generated by integrating out high energy $(b.f)$ modes.
For energy scales well below $T_b$ the boson has a well defined amplitude, and we may write
$b_r \simeq b_0 e^{i\theta_r}$
with $b_0 \sim \sqrt{x}$.
The phase $\theta_r$ is `Higgsed' and is described by the Gaussian action
\be
S_{eff}[\theta,a] = \int_{\tau} \sum_r \frac{\kappa_b}{2}\left(\partial_\tau \theta_r - a_{0r} \right)^2 + \sum_{rr'} \frac{\rho_{bs}}{2}\left(\nabla_i \theta - a_i \right)^2
\ee
where $\kappa_b, \rho_{bs} \sim x$ are the compressibility and phase stiffness of the boson. The index $i$ refers to the two spatial directions, and $\nabla_i$ is a lattice derivative. Now we shift $a_\mu \rightarrow a_\mu - \partial_\mu \theta$ and introduce the fermion field
\be
q_{r\alpha} = e^{-i\theta_r}f_{r\alpha}
\ee
An effective action, appropriate below the scale $T_b$, then becomes
\bea
\label{IFLS}
S_{IFL}[q,a] & = & \int_\tau \sum_r \bar{q}_{r\alpha} \left(\partial_\tau - ia_0 - \mu \right) q_{r\alpha} \\
& & - \int_{\tau} \sum_{<rr'>} t_s \left(\bar{q}_{r\alpha}q_{r'\alpha} e^{ia_{rr'}} + h.c \right) \\
& & + \int_\tau \sum_r \frac{\kappa_b}{2}  a_{0r}^2 + \sum_{rr'} \frac{\rho_{bs}}{2} \vec a^2
\eea

The operator $q_{r\alpha}$ is gauge invariant and has the same quantum numbers as the electron operator. Specifically
$c_{r\alpha} = b_0 q_{r\alpha}$.
Thus the true electron spectral function differs from that of the $q$-field by an overall factor of $b_0^2 \sim x$.
The action $S_{IFL}[q,a]$ represents electron-like quasiparticles coupled to a {\em massive} gauge field. It can be understood as a theory of electrons coupled to strong density-density and current-current interactions, as is readily seen by integrating out the gauge field. Thus despite our derivation using spin-charge separated variables the IFL regime is {\em not} spin-charge separated.
As the coefficients $\kappa_b, \rho_{bs} \sim x$ the resultant four fermion interactions have strength of order $1/x$ and hence are strong.

We proceed with a standard RPA treatment to determine the effective gauge action by integrating out the $q_\alpha$. The $a_0$ component receives a large $x$-independent constant contribution coming from the bulk compressibility of the fermions which represents the usual Debye screening. A common approximation is to drop the terms involving $a_0$, and focus on the transverse component $a_i$. It is however more correct to retain it as an effective $x$-dependent on-site Hubbard repulsion $U_{eff}(x)$. For not too small doping this repulsion will not change the qualitative physics but will play a role in determining things like the superconducting $T_c$. The more dangerous transverse gauge field has an effective action
\be
\label{gaugeS}
S_{eff}[a] = \frac{1}{2\beta}\sum_{\omega_n}\int_{\vec p} \left(\Pi(\vec p, i\omega_n)  +   \rho_{bs} \right)|a(\vec p, i\omega_n)|^2
\ee
$\Pi$, the current-current polarizability of the $q$-fermions, is given (after continuing to real frequencies $i\omega_n \rightarrow \omega + i0^+$)   by
\bea
\label{pol}
\Pi(\vec p, \omega)& = & i\omega \sigma(p, T)+ \chi p^2 \\
\sigma(k, T) & = & \frac{v_F k_0}{\sqrt{4\gamma_{tr}^2 + v_F^2 p^2}}
\eea
Here $\sigma(0,T)$ may be interpreted as the conductivity of the $q$-fermions, and $\gamma_t$ is an appropriate (transport) scattering rate. $v_F$ is the Fermi velocity and the coefficient
$k_0 \sim K_f$ (the Fermi wave vector). The diamagnetic susceptibility $\chi \sim t_s d^2 \sim \frac{1}{m}$ is of order the inverse fermion mass ($d$ is the lattice spacing).  F or $T \ll T_b$ the scattering rate $\gamma_{tr}$ found below is sufficiently small that it can be ignored compared to the typical momentum $p \sim \sqrt{\frac{\rho_{bs}}{\chi}}$.  Then the typical frequency of a gauge fluctuation of momentum $p$ scales as $\omega \sim \frac{\chi p^3}{k_0}$. The transverse gauge field $\vec a$ thus gets a gap of order
\be
\label{Tcoh}
T_{coh}  \sim  \rho_{bs} \sqrt{\frac{\rho_{bs}}{E_F}}
\ee
where $E_F$ is the Fermi energy. This gap, which is parametrically smaller than $T_b \sim \rho_{bs}$, sets a scale above which the gauge field has still not realized that the boson has condensed.

At intermediate temperatures $T_b \gg T \gg T_{coh}$ the $q$-field is strongly scattered by the `gapless' gauge fluctuations. It is only on cooling below $T_{coh}$ does the gauge field become truly gapped, and the scattering subsides to produce a Landau fermi liquid. Thus the coherence scale  is given by Eqn. \ref{Tcoh} and not by  $T_b \sim x$ as previously assumed.

Despite being metallic, IFL has the same spin physics as the insulating spin liquid with a spinon Fermi surface. Thus it will have a constant spin susceptibility, and enhanced ``$2K_F$" spin correlations\cite{aliomil} ({\em i.e} at wavevectors connecting tangential portions of the Fermi surface). Further at low temperature (but still above $T_{coh}$), the electronic specific heat will have a $T^{2/3}$ temperature dependence exactly as in the corresponding insulating spin liquid. Interestingly as the $q_\alpha$ field carries both spin and charge the enhanced $2K_F$ correlations will also show up in the charge response.

Consider the single particle Greens function. The $q$-field acquires a non-trivial self energy due to scattering off thermally excited low frequency gauge fluctuations.  In leading order perturbation theory,  the scattering rate $\gamma$ at a Fermi surface point $\bK$ is determined from the imaginary part of the $q$ self energy $\Sigma''(\bK, \omega = 0)$ and is given by
\bea
\label{gammapert}
 \gamma & = &  \pi  \int_{\bp,\Omega} \left(\textbf{v}_\textbf{F} \times \hat{\textbf{p}}\right)^2
 D^{''}(\textbf{p},\Omega)A\left(\bK - \textbf{p}, - \Omega \right) \nonumber \\
 && \left(n(\Omega) + f(\Omega)\right) \nonumber
\eea
Here $\textbf{v}_\textbf{F}$ is the Fermi velocity, $D''$ is the spectral density of the gauge field propagator, and $A$ is the spectral function of the $q$-field. $n(\Omega), f(\Omega)$ are the Bose and Fermi functions respectively. We use
$A\left(\bK - \textbf{p}, - \Omega \right)  =  \delta(-\Omega -\epsilon_{\bK - \textbf{p}})$
where $\epsilon_{\bp}$ is the dispersion of the $q$-field. For $\bK$ at the Fermi surface,
$\epsilon_{\bK - \textbf{p}}$ may be expanded as $-v_F p_\parallel + c p_\perp^2$ with $p_\parallel, p_\perp$ the two components of $\bp$ parallel and perpendicular to the normal to the Fermi surface. The constant $c$ is of order the inverse quasiparticle mass. This fermion dispersion implies that  in the self energy integral the typical $|p_\parallel| \sim p_\perp^2 \ll |p_\perp|$. Thus we may drop $p_\parallel$ in the gauge propagator and do the
$p_\parallel$ integral to get
\be
\label{gammapert1}
\gamma  =  \pi v_F  \int_{\Omega}\left(\frac{1}{sinh(\beta \Omega)}\right)\int_0^\infty dp D^{''}(p, \Omega)
\ee
The gauge propagator is obtained from Eqns. \ref{gaugeS} and \ref{pol}, and replacing the polarizability by its zero temperature form (justified as the scale for the momentum $p$ is the inverse gauge field screening length $\sqrt{2m\rho_{bs}} \gg \gamma_{tr}/v_F$).
We may then write the result of the $p$-integration in the following form
\be
 \gamma =  \frac{2\pi v_F k_0}{\rho_{bs}^2}\int_0^\infty d\Omega \frac{\Omega}{sinh(\beta \Omega)}g\left(\frac{\Omega}{T_{coh}} \right)
\ee
where
\be
g(t) = \int_0^\infty dx \frac{x}{t^2 + \left(x^3 + x\right)^2}
\ee
This has the limiting behaviors $g(t \rightarrow 0) \sim \ln\frac{1}{t}$, $g(t \rightarrow \infty) \sim \frac{1}{t^{\frac{4}{3}}}$. In the IFL regime we have $T \gg  T_{coh}$. The $\Omega$ integral can be done in this limit and we find a linear $T$ single particle scattering rate
\be
\label{gamma}
\gamma \simeq \left(k_B T \right) \sqrt{\frac{E_F}{\rho_{bs}}}
\ee



As the scattering off the $a$ fluctuations is small angle the transport scattering rate is potentially different from the single particle one. Accordingly Eqn. \ref{gammapert} must be modified by an additional factor of $\frac{p_\perp^2}{p_F^2}$.  We then find
a transport scattering rate $\gamma_{tr} \sim T^{\frac{4}{3}}$ (and hence a $T^{\frac{4}{3}}$ resistivity) at the lowest temperatures. The same rate was found\cite{leenag} for spinon transport in the NFL regime  but is now interpreted as the transport rate of electrons in IFL.

Next we consider the underdoped region. The mean field calculation\cite{kotliarliu} demonstrates that below a temperature scale $T^*$ a non-zero $d_{x^2 - y^2}$
pair amplitude develops. Within the {\em mean field theory} the superconducting transition happens at a $T_c = min(T_b, T_\Delta)$. Beyond mean field theory, $T_c$ will be suppressed by various fluctuation effects, as we discuss at the end of the paper.
Thus for moderate underdoping the pseudogap region (PG of Fig. \ref{tJslvpdia}b) must be described as IFL modified by the presence of a local $d$-wave pair amplitude but no global phase coherence. We now develop a description of the single particle properties in the PG regime.

 In the PG regime the action Eqn. \ref{IFLS} must be supplemented by the coupling to a fluctuating $d$-wave pair field $\Delta(r, \tau)$. We write
\bea
S[q, \Delta, a] & = & S_{IFL} + S[\Delta, q] + S[\Delta, a] \\
S[\Delta, q] & = & \int_{\tau} \sum_r \Delta^*\sum_{r' \in r} \eta_{rr'} \left(q_{r\uparrow}q_{r'\downarrow} - q_{r\downarrow}q_{r'\uparrow}\right) \\
& & + c.c
\eea
Here $\eta_{rr'} = +1$ on horizontal bonds and $-1$ on vertical bonds. We have thus taken $\Delta$ to couple to a $d$-wave singlet pair of the $q$-fields that has center of mass at the site $r$. We have not explicitly written out the term $S[\Delta, a]$ which describes the action for the boson field $\Delta$ coupled minimally with charge $2$ to the $a$-field. In a finite temperature regime in which $\Delta$ has a well-defined amplitude $\Delta_0$ but fluctuating phase  its correlators   will have the form
\be
\langle \Delta^*(r,t) \Delta(0,0) \rangle = \Delta_0^2
F(r, t)
\ee
such that
$F(0, 0)  =  1$ while for  $|t| \gg \frac{1}{\Gamma}$, or $|r| \gg \xi_{\phi} \sim v_F/\Gamma$,  $F(r,t)$ decays exponentially to zero. $\Gamma$ is thus the Cooper pair phase decay rate.

Previous work\cite{norman,chubukov,tsleehitcphen} showed how the combination of a linear -$T$ scattering rate together with the local $d$-wave pair amplitude lead directly to an electron spectral function that has an antinodal pseudogap coexisting with gapless $T$-dependent
Fermi arcs near the nodal direction. Indeed if Eqn. \ref{gamma} survives the development of the pair amplitude then the electron self energy has an extra contribution
\be
\Sigma_p (\bK, \omega) \simeq \frac{\Delta_{0\bK}^2}{\omega + \epsilon_\bK + i \gamma}
\ee
which was used to successfully fit the ARPES data\cite{norman}.

The key microscopic question therefore is to justify Eqn. \ref{gamma} even in the presence of a local pair amplitude. We now show that the assumption of such a scattering rate is self-consistent
within the slave particle gauge theory.
To understand this we first need the form of the gauge propagator in this pseudogap region. The polarizability in
Eqn. \ref{pol} will now get contributions from both the $q$ and $\Delta$ fields. These may be described in terms of a {\em frequency} dependent function
\be
\label{sigma}
\sigma(k,\omega, T) \simeq \frac{K(\omega)}{i\omega} + \frac{L(T) k_0 v_F}{\sqrt{4\gamma_{tr}^2 + v_F^2 k^2}}
\ee
The first term is the pair field contribution with a frequency dependent phase stiffness $K(\omega)$
satisfying $K(\omega) \approx 0$ for $|\omega| \ll \Gamma$ and
$K(\omega) \approx \rho_{fs}$ (the `bare' phase stiffness of the $\Delta$ field) for $\Gamma \ll |\omega| \ll \Delta_0$.
 The second term in Eqn. \ref{sigma} is the ``quasiparticle" contribution coming from the gapless Fermi arc region. Its coefficient is reduced by a factor $L(T) \sim \frac{T}{T^*}$ which reflects the fractional length of the gapless arc (in agreement with the experiments of Ref. \onlinecite{kanigel}). For moderately underdoped samples $L(T)$ is a sizeable fraction even close to $T_c$.

The scattering rate of the $q$-field at a Fermi point in the arc region off these gauge fluctuation is given by Eqn. \ref{gammapert} but with modified gauge and electron spectral functions. Recognizing that the electron spectrum in the arc region $\gamma \gg |\Delta_{0K}|$ is little modified from that in IFL, the $\vec p$ integrals may be done as before. The important contribution again comes from $|\Omega| \ll T$ so that
\be
\gamma \sim Tv_F k_0 L \int_0^\infty \frac{d\Omega}{\left(\rho_{bs} + K(\Omega)\right)^2}g\left(L \Omega\sqrt{\frac{E_F}{\left(\rho_{bs} + K(\Omega)\right)^3}}\right)
\ee
The function $g$ has most of its support in a region where its argument is $< o(1)$. Thus so long as
$\Gamma \gg \frac{\rho_{bs}}{L}\sqrt{\frac{\rho_{bs}}{E_F}} \sim \frac{T_{coh}}{L}$ the important frequency range has $K(\Omega) = 0$. Then the scattering rate continues to be given by Eqn. \ref{gamma}. If $\Gamma$ falls below this scale the scattering rate will drop as the relevant low frequency gauge fluctuations become stiffer due to the contribution from the pair field.

As  $T$ is increased from $T_c$, we expect that $\Gamma$ drops rapidly (at least in the critical region within Kosterlitz-Thouless theory). Indeed the experiments of Ref. \onlinecite{corson} show that $\Gamma \sim T$ already about $20 K$ above $T_c$. (However $\Gamma \sim \Delta_0$ only at much higher temperatures.) Consequently for moderately underdoped samples the condition for incoherent single particle scattering will be satisfied all the way down to within a narrow temperature range of the superconducting transition. Thus our original assumption of a large $\gamma$ is self-consistent almost all the way upto $T_c$, and this implies the
the emergence of both the pseudogap and the gapless $T$-dependent Fermi arcs.


We emphasize that this mechanism for the incoherent scattering in the normal state is distinct from previous slave particle theories where the $T_c$ was tied to the bose condensation of the holons. Rather here it is the freeing up of the low frequency gauge fluctuations by crossing $T_c$ that leads to the incoherent scattering.
Clearly in our theory coherent quasiparticles always exist at $T = 0$  so that a magnetic field sufficient to suppress $T_c$ is not sufficient to suppress the quasiparticle coherence scale, as assumed in Ref. \onlinecite{tsleehitcphen}.

Finally we comment on the effects of the on-site effective Hubbard repulsion $U_{eff(x)}$ which should increase with decreasing $x$. First this will lead to short ranged antiferromagnetic spin correlations, and eventually to antiferromagnetism for very small $x$. Second, $U_{eff}$
will lead to a suppression of the superfluid stiffness $\rho_{sf}$ of the pair field $\Delta$ from its BCS value. This in turn will bring down the true superconducting transition temperature $T_c$ from the pairing scale $T^*$. A different mechanism for the suppression of $T_c$ is due to the thermal excitation of current carrying nodal quasiparticle states proposed in Ref. \onlinecite{leewen}. Within the present theory the current carried by such a quasiparticle is proportional to $x$ for small $x$. However numerical calculations\cite{nave} with projected wave functions show that for moderate $x$ the current rises rapidly. As our theory anyway is designed only for such doping levels this may be a viable mechanism. We leave for the future a more elaborate theory of these effects.

In summary we explored issues of single particle coherence and pairing within the slave boson gauge theory of the doped Mott insulator. We studied a new non-fermi liquid regime, characterized by a linear-$T$ single particle scattering rate, which extends to much lower temperature than the boson condensation scale. In the underdoped side this state acquires a $d$-wave pair amplitude below a $T^*$ line, and eventually becomes a $d$-wave superconductor. In the regime between $T_c$ and $T^*$, there is a pseudogap in the single particle spectrum which coexists with `gapless' Fermi arcs near the nodal direction that shrink as the temperature is reduced. Thus this model captures many of the most striking phenomenological aspects of cuprate physics. At present we however do not have a natural explanation of the linear resistivity that is seen at temperatures above $T^*$. Perhaps this difficulty can be overcome in the future.

   TS was supported by NSF Grant DMR-0705255, and PAL by NSF Grant DMR-0804040.

\end{document}